\documentclass[journal]{IEEEtran}
\usepackage{mathrsfs}
\usepackage{amsmath}
\usepackage{bm}
\usepackage[cmintegrals]{newtxmath}
\usepackage{threeparttable}
\usepackage{longtable}
\usepackage{makecell}
\usepackage{multirow}
\usepackage{color}
\usepackage{booktabs}

\usepackage{amsfonts,amssymb}

\ifCLASSINFOpdf
  \usepackage[pdftex]{graphicx}
  \graphicspath{{../pdf/}{../jpeg/}}
\else
\fi
\begin{document}
%
\title{Visual Perception Model for Rapid and Adaptive Low-light Image Enhancement}
%
%
%

\author{Xiaoxiao~Li,
        Xiaopeng~Guo,
        Liye~Mei,
        Mingyu~Shang,
        Jie~Gao,
        Maojing~Shu,
        and Xiang~Wang
        
\thanks{X. Li is with the School of Information Science and Engineering, Yunnan University, Kunming 650091, China (email: lxxwxy@outlook.com). (\emph{Corresponding author: Xiaopeng~Guo.})}
\thanks{X. Guo, M. Shang, J. Gao and M. Shu are with the Wangxuan Institute of Computer Technology, Peking University, Beijing 100080, China (email: xpguo123@gmail.com; mingyu.shang@pku.edu.cn; gaojie2018@pku.edu.cn; shumaojing@pku.edu.cn).}
\thanks{L. Mei is with the Institute of Technological Sciences, Wuhan University, Wuhan 430072, China (email: liye\_mei@outlook.com).}
\thanks{X. Wang is with the Department of Electronic Engineering, Tsinghua University, Beijing 100084, China (email: airablewang@163.com).}}

%
%

\markboth{}%
{Shell \MakeLowercase{\textit{et al.}}: Bare Demo of IEEEtran.cls for IEEE Journals}
%



\maketitle

\begin{abstract}
Low-light image enhancement is a promising solution to tackle the problem of insufficient sensitivity of human vision system (HVS) to perceive information in low light environments. Previous Retinex-based works always accomplish enhancement task by estimating light intensity. Unfortunately, single light intensity modelling is hard to accurately simulate visual perception information, leading to the problems of imbalanced visual photosensitivity and weak adaptivity. To solve these problems, we explore the precise relationship between light source and visual perception and then propose the visual perception (VP) model to acquire a precise mathematical description of visual perception. The core of VP model is to decompose the light source into light intensity and light spatial distribution to describe the perception process of HVS, offering refinement estimation of illumination and reflectance. To reduce complexity of the estimation process, we introduce the rapid and adaptive $\bm{\beta}$ and $\bm{\gamma}$ functions to build an illumination and reflectance estimation scheme. Finally, we present a optimal determination strategy, consisting of a \emph{cycle operation} and a \emph{comparator}. Specifically, the \emph{comparator} is responsible for determining the optimal enhancement results from multiple enhanced results through implementing the \emph{cycle operation}. By coordinating the proposed VP model, illumination and reflectance estimation scheme, and the optimal determination strategy, we propose a rapid and adaptive framework for low-light image enhancement. Extensive experiment results demenstrate that the proposed method achieves better performance in terms of visual comparison, quantitative assessment, and computational efficiency, compared with the currently state-of-the-arts. A MATLAB implementation will be provided to facilitate the future research of low-light image enhancement field. 
\end{abstract}


%
\IEEEpeerreviewmaketitle

\section{Introduction}
%
%
%
%
\IEEEPARstart{H}{uman} 
 vision system (HVS) is always influenced by ambient light. Images captured in low-light environment always have less detail and low contrast, which is difficult for HVS to perceive \cite{yu2017low}, \cite{awad2019adaptive}; also, the low-light image could seriously affect other computer vision tasks that highly rely on target visibility, including saliency detection \cite{yuan2016dense}, semantic segmentation \cite{long2015fully}, and object tracking \cite{zhang2015graph}, etc. Therefore, in order to meet the requirements of both the good perception of HVS and the promotion of other vision tasks, low-light image enhancement technique is definitely necessary.

Over the decades, researchers have proposed various methods to boost the performance of low-light image enhancement. In generally, these methods could be roughly classified into three categories: \emph{pixel-based} method, \emph{model-based} method and \emph{learning-based} method.


\begin{figure}[t]
	\centering
	\includegraphics[width=0.48\textwidth]{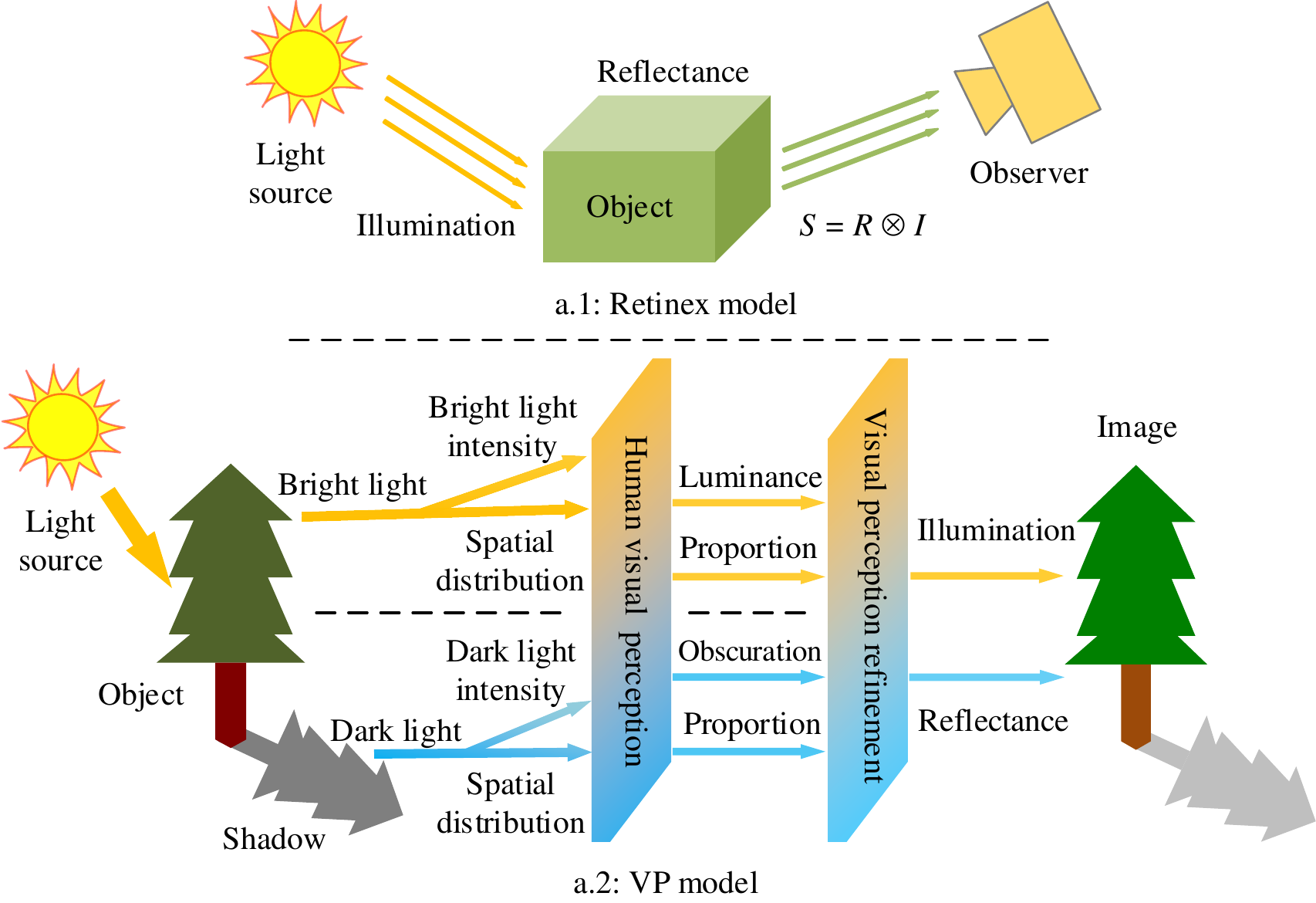}
	\caption{The comparsion of modeling low-light image by the proposed VP model and the classical Retinex model.}
	\vspace{-0.4cm}
\end{figure}
\begin{figure*}[t]
	\centering
	\includegraphics[width=0.8\textwidth]{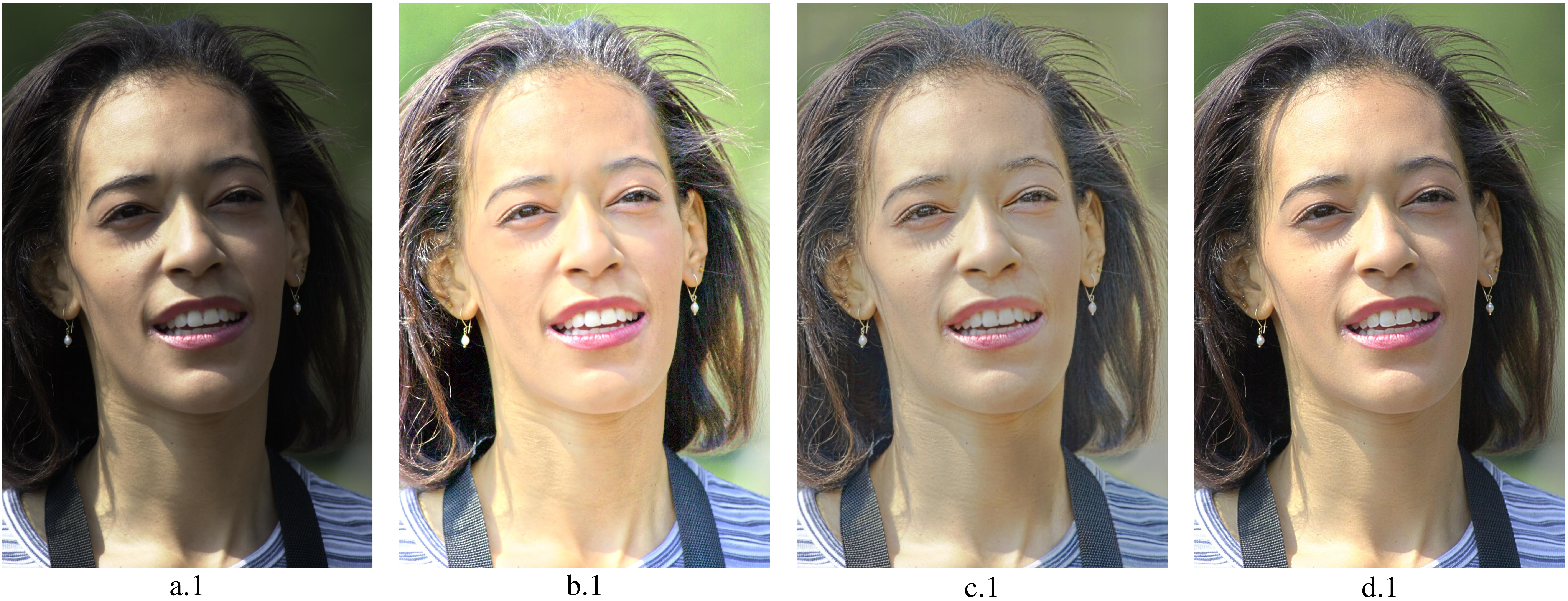}
	\caption{(a.1): a challenging uneven exposure image; (b.1-c.1): enhanced results by state-of-the-art methods [25] and [29]; d.1: enhanced result by our method.}
\end{figure*}
In terms of early pixel-based methods, illumination amplification is the most intuitive method. However, this method may cause the over-saturation problem, image details are inevitably lost. Histogram equalization \cite{pizer1987adaptive}, \cite{han2010novel} could effectively alleviate this problem by adjusting the dynamic range. Besides, the enhancement methods based on the intensity mapping function also have been proposed \cite{bennett2005video}, \cite{shan2009globally}, \cite{yuan2012automatic}. Unfortunately, These early attempts are difficult to visualize the underlying structure of the dark areas due to the lack of illumination estimation. From different perspectives, some researchers  \cite{dong2011fast}, \cite{zhang2012enhancement}, \cite{li2015low} attempt to find the eccentric similarity between inverted image and fuzzy image. For instance, the dehazing-based method \cite{li2018structure} finishes enhancement task via establishing the inverse links. 

Model-based methods are proposed by establishing the solid theoretical foundation. There are copious literature about this method \cite{yu2017low}, \cite{kimmel2003variational}, \cite{ng2011total}, \cite{wang2013naturalness}, \cite{fu2016weighted}, \cite{park2017low}, \cite{cai2017joint}. The classic Retinex model \cite{land1977retinex} abstracts color images into illumination and reflectance by mapping real physical environments. Subsequent studies generally simplify Retinex model by logarithmic transformation \cite{jobson1997multiscale}. However, this kind of method is hard to obtain stable illumination intensity, resulting in unnatural image details \cite{fu2016weighted}. To solve this problem, Wang \emph{et al.} \cite{wang2013naturalness} propose a naturalness preserving enhancement (NPE) method for uneven illumination image. Furthermore, Guo \emph{et al.} \cite{guo2016lime2} introduce a low-light image enhancement method (LIME), which reconstructs the contrast of illumination by selecting the maximum intensity of each channel. However, the enhanced results of these methods always produce noise. Hence, in \cite{li2018structure}, a robust Retinex model (RRM) is designed to try to reduce the undesirable noise. Also, some researchers attempt to build other models to obtain better enhancement results. For instance, Ying \emph{et al.} \cite{ying2017new} propose a nonlinear camera response model (CRM) to estimate the image exposure. Wang \emph{et al.} \cite{wang2019low} establish the absorption light scattering model (ALSM) by explaining the absorption light imaging process. Some more sophisticated models \cite{wang2019low}, \cite{ying2017new} with high computational complexity also have been proposed. Though these model-based methods achieve inspiring enhancement results, they model the real environment from simple perspective, limiting their performance to some extent; also, the model-based methods always own many empirical parameters, such as \cite{wang2019low}, \cite{ying2017new}, reducing the models' adaptive ability.
\begin{table*}[t]
	\centering
	\setlength{\tabcolsep}{1.5mm}{
		\caption{The Notations and Terms Used in This Paper.}	
		\begin{tabular}{|l|l|}
			\hline
			$S$, $F_o$ & Input image, output image, for respectively. \\
			$I_c$, $I_d$, $I_b$ & Brightness channel image, detail image with pseudocolor, smooth image, for respectively. \\
			$H$, $H_b$, $H_d$ & Image pixel energy with pseudocolor, bright areas, dark areas, for respectively.\\
			$\bm{\phi}_1$, $\bm{\phi}_2$, $\bm{\varphi}_1$, $\bm{\varphi}_2$ & Luminance, obscuration, spatial distribution of luminance, spatial distribution of obscuration, for respectively. \\
			$P_1$, $P_2$, $Q_1$, $Q_2$ & Bright energy, dark energy, the area ratio of bright, the area ratio of dark, for respectively.  \\		
			$\bm{\beta}$, $\bm{\gamma}$ & The propsoed two adaptive functions from the quantification of visual information.\\
			$M_e$, $N_e$, $I_e$, $F_e$ & Estimated illumination, estimated reflectance, reconstructed brightness channel image, intermediate image, for respectively.\\
			$K$, $T$ & The number of cycle operations, adaptive threshold, for respectively.\\
			\hline
	\end{tabular}}
\end{table*}
\begin{figure*}[t]
	\setlength{\abovecaptionskip}{-0.1cm}
	\setlength{\belowcaptionskip}{-0.5cm}
	\centering
	\includegraphics[width=1\textwidth]{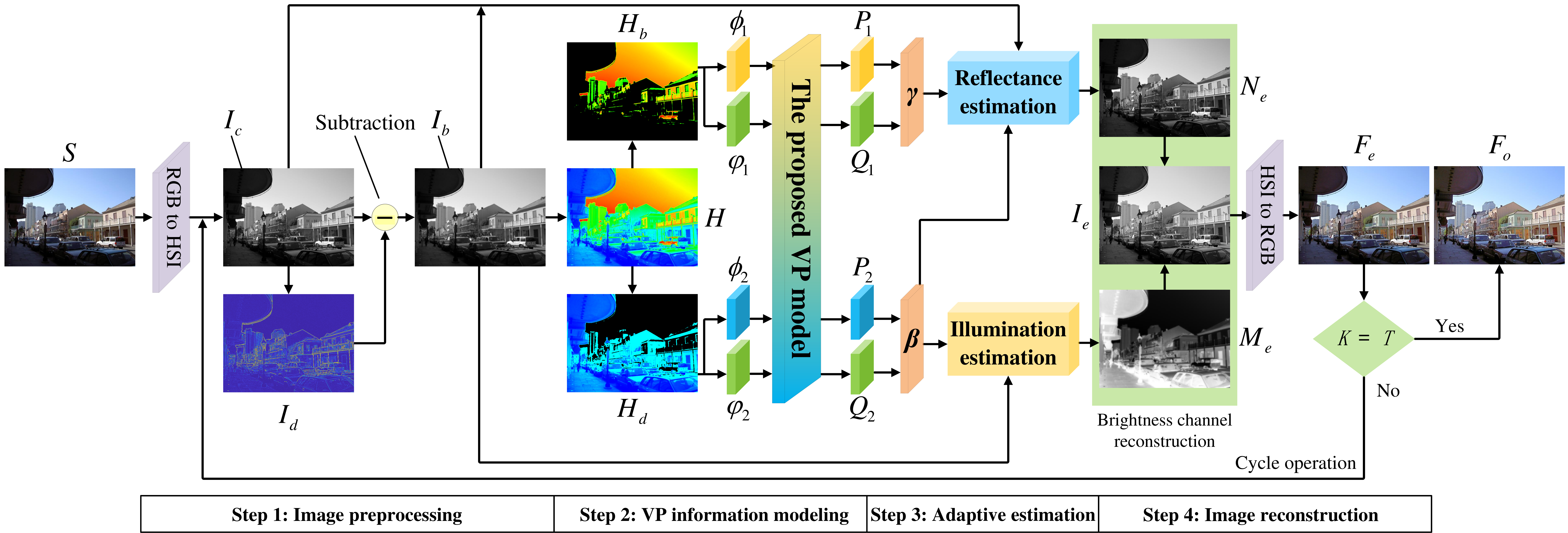}
	\caption{The proposed VP-based framework for low-light image enhancement. Our framework is implemented in four steps: first, we preprocesses the brightness channel to obtain a smooth image $I_b$ to improve the accuracy of modeling visual perception information; then, we utilize the proposed VP model to decompose the visual perception features of images to establish the precise relationship between images and visual perception; third, we construct $\bm\beta$ and $\bm\gamma$ functions as estimation scheme to rapidly and adaptively adjust the intensity of illumination and reflectance estimation; finally, we give a optimal determination strategy to reconstruct the enhanced image. More details of the proposed framework will be provided in Section III-A.}
	\vspace{-0.4cm}
\end{figure*}

In recent years, with the rapid development of deep learning \cite{lecun2015deep}, there emerge some learning-based methods \cite{yan2016automatic}, \cite{gharbi2017deep}, \cite{zhang2019kindling}, \cite{chen2018learning} that apply deep learning technniques to model the enhancement task. Gharbi \emph{et al.} \cite{gharbi2017deep} implement real-time image enhancement by embedding bilateral grid processing \cite{chen2007real} into the neural network. Chen \emph{et al.} \cite{chen2018learning} develop a end-to-end pipeline for enhancing low-light images. Inspired by Retinex theory, Zhang \emph{et al.} \cite{zhang2019kindling} build a practical low-light image enhancer from multi-exposure images, named “kindling the darkness" (KIND). Ren \emph{et al.} \cite{ren2019low} present a deep hybrid network to enhance images by jointly learning salient structures and global content. Although the performance of these learning-based methods are impressive, most of them require large-scale training datasets with ground truth. It's time-consuming to establish such a training datset to provide supervision learning. Particularly, in extremely low light conditions, the task of collecting a large amount of clear normal images becomes harder.

To address the problems above, we propose visual perception (VP) model tailored to low-light image enhancement in this paper. VP model describes rich light properties to analyze visual perception of the imaging environment. Unlike the classical Retinex model directly abstracts light into illumination and reflectance (show in Fig.1-a.1), we model the visual perception information (see in Fig.1-a.2) from light source to establish a precise relationship between visual perception and image feature information to automatically adjust the enhancement intensity. 

More specifically, considering the light in bright or dark environment may cause visual photosensitive imbalances (VPI), we intend to estimate enhancement weights from two productive ways: 1) we analyze the bright/dark level and area ratio of low-light images corresponding to the light intensity and light spatial distribution to model the visual perception information; 2)  by quantifying visual perception information to describe the degree of VPI, we determine the enhancement weights of bright and dark areas and realize the refinement of illumination and reflectance estimation. it's worth noting that VP model contains much parameters, bringing high computational complexity. We therefore provide rapid and adaptive $\bm{\beta}$ and $\bm{\gamma}$ functions to build illumination and reflectance estimation schemes. $\bm{\beta}$ and $\bm{\gamma}$ functions quantify visual perception information to respectively define the degree of VPI in dark and bright areas, realizing adaptive estimation of enhancement weights; also, $\bm{\beta}$ and $\bm{\gamma}$ functions integrate the parameters (see in Fig.3) of VP model to reduce the computational complexity and thus obtain the ability of rapid enhancement. Furthermroe, we employ \emph{cycle operation} and \emph{comparator} to establish a optimal determination strategy to further enhance the visual perception consistency of the enhanced results. Concretely, the \emph{comparator} is responsible for determining the “\emph{optimal}” enhancement results from multiple enhanced results through implementing the \emph{cycle operation}. This configuration further improves the adaptivity. 

Through the proposed VP model, illumination and reflectance estimation scheme and optimal determination strategy, we establish rapid and adaptive framework for low-light image enhancement. Fig.2 illustrates that our framework exhibits better enhancement results, compared with other state-of-the-art methods. We believe that the proposed VP model is a promising solution for low-light image enhancement and will encourage future research. To sum up, the main contributions of this work could be summarized as follows:

\begin{itemize}
	\item We propose a novel visual perception (VP). First, different from the Retinex-based model that abstracts light illumination and reflectance roughly, we go further and refine light into light intensity and light spatial distribution to establish the precise relationship between HVS and image feature information, aiming at adjusting the enhancement intensity automatically; second, to further boost the visual perception of the enhanced result, joint illumination and reflection estimation are given. 
\end{itemize}
\begin{itemize}
	\item We introduce an adaptive scheme for estimating illumination and reflectance. Two elaborate functions, $\bm{\beta}$ and $\bm{\gamma}$ are designed to enhance the intensity of dark and bright areas, pushing adaptivity performance further; also, the empirical parameters are thus reduced greatly, achieving a rapid enhancement process.
\end{itemize}
\begin{itemize}
	\item To obtain the “\emph{optimal}” enhancement results, we give a optimal determination strategy via employing \emph{cycle operation} and \emph{comparator}. The \emph{comparator} is responsible for determining the optimal enhancement results from multiple enhanced results through implementing the \emph{cycle operation}. This effective strategy further improves the adaptive ability of enhancing low-light images.
\end{itemize}
\begin{itemize}
	\item By leveraging the proposed VP model, adaptive scheme, and optimal determination strategy together, an adaptive and rapid framework for low-light image enhancement is given. Experimental results demonstrate that the proposed framework is superior than the state-of-the-art methods not only on visual perception, but also quantitative assessment. Besides, the higher computational efficiency of our framework is also verified.
\end{itemize}

The rest of this paper is organized as follows. Motivation of this work is given in Section II. In Section III, we present our framework in detail. Experimental discussions are provided in Section IV. Finally, Section V concludes the paper.

\section{Motivation}
The classical Retinex theory divides the observed image into illumination and reflectance (in Fig.1-a.1) to simulate the HVS, which can be formulated as:
\begin{eqnarray}
S &=& R \otimes I,
\end{eqnarray}
where $S$ is the observed image, $I$ indicates illumination, $R$ represents reflectance, and the operator $\otimes$ denotes element-wise multiplication.

The illumination represents the light received by object. It generally shows ambient light into bright/dark state. Reflectance represents the intrinsic properties of object, which is thought to be consistent under any light condition. Based on this view, some researchers \cite{wei2018deep}, \cite{yue2017contrast} focus on decomposing reflectance images. However, mere reflectance is difficult to separate because of the interference from illumination. Also, it's well known that there are always distinctiveness in visual perception between low-light images and natural images. Nevertheless, Retinex model could not uses the human sensory information to estimate natural images, those methods based on illumination/reflectance estimation thus always lack sufficient adaptability and the enhanced results are short of detailed visual perception. By contrast, we try to rethink the Retinex model in an distinctive way. That's to say, we propose to use estimated visual perception distinctiveness to adaptively enhance low-light images rather than depth-based decomposition of illumination and reflection. Hence, we refine light into light intensity and light spatial distribution to establish the precise relationship between HVS and image feature information. The VP model is thus designed to model the visual perception of HVS. Our VP model has a better representation capacity than Retinex model in the perception of image visual information, providing a promising solution for rapid and adaptive enhancement of low-light images.

\begin{figure}[t]
	\centering
	\includegraphics[width=1\linewidth]{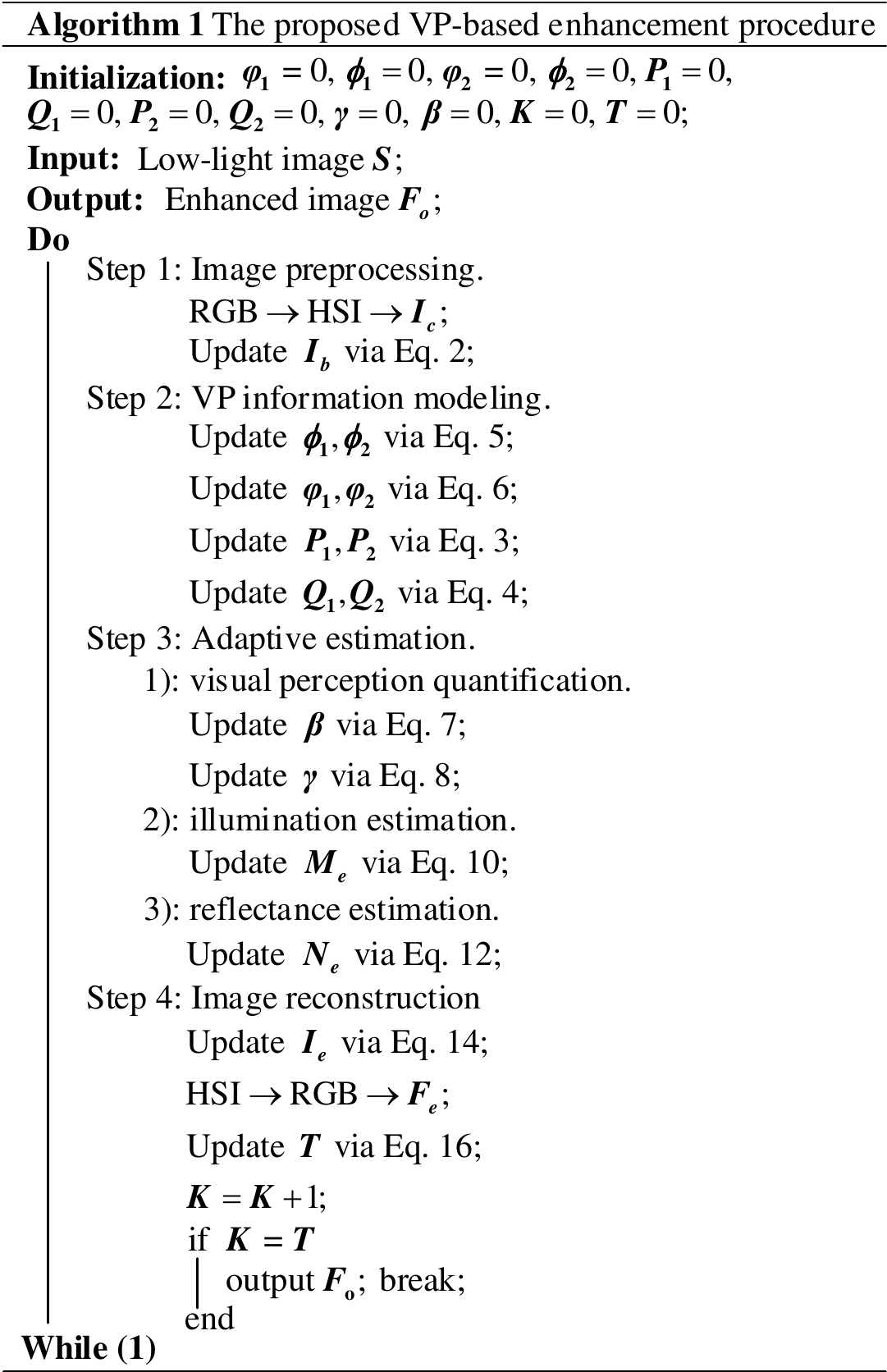}
\end{figure} 
\section{Methodology}
\subsection{Overview}
We present a comprehensive overview of the proposed framework in the section. For the convenience of description, we summarize the commonly used notations in TABLE I. We then build the proposed framework from four steps. \textbf{Step 1}: we extract the brightness channel of the input image by HSI color space. Whereafter, we pretreat the brightness channel by subtracting the detail image $I_d$ from $I_c$. \textbf{Step 2}: we use the image $H$ intuitively describe the VP interesting information: $\bm{\phi}_1$ (luminance), $\bm{\varphi}_1$ (spatial distribution of luminance), $\bm{\phi}_2$ (obscuration), and $\bm{\varphi}_2$ (spatial distribution of obscuration). The interesting information is quantified into corresponding digital quantities: $P_1$ (bright energy), $Q_1$ (the area ratio of bright), $P_2$ (dark energy), and $Q_2$ (the area ratio of dark). Then, we integrate these digital quantities into $\bm{\beta}$ and $\bm{\gamma}$ to estimate the degree of VPI. \textbf{Step 3}: we use adaptive $\bm{\beta}$ and $\bm{\gamma}$ to estimate illumination $M_e$ and reflectance $N_e$. \textbf{Step 4}: the estimated illumination and reflectance are used to reconstruct the new brightness channel $I_e$, then we set up a comparator to select the final output image $F_o$ from the intermediate image $F_e$. In detail, we summarize the main steps of the proposed method in \textbf{Algorithm 1}.

\subsection{Image Preprocessing}
We obtain the brightness channel $I_c$ by converting input image to the HSI space in Fig.3. We use average filter to extract the details of the $I_c$. The image details are the potential interference signal for VP modeling interesting information since it contains the height-variability pixel gradient; on the contrary, the pixels in the boundary areas of the bright/dark usually change gradually. So $I_b$ as an input signal has a higher precision than $I_c$ for decomposing bright/dark areas. In addition, the $I_b$ could effectively improve the structure information of the new brightness channel $I_e$. Because $I_b$ is the signal source of illumination estimation (in Fig.3), we consider the subtractive detail as the dark areas, which gives the detail a larger weights by VP. In summary, the pretreatment process has two-fold advantages: improving the accuracy of interesting information modeling by VP and enhancing the structure information of low-light image. The pretreatment process can be formulated as:
\begin{eqnarray}
I_b &=& \left \{ I_d=I_c-\lambda I_c, I_d > 0 \left | \lambda I_c - I_d \right . \right \},
\end{eqnarray}
where $I_d$ is the detail image, and $\lambda$ represents the average filter. 
\begin{figure}[t]
	\centering
	\includegraphics[width=0.38\textwidth]{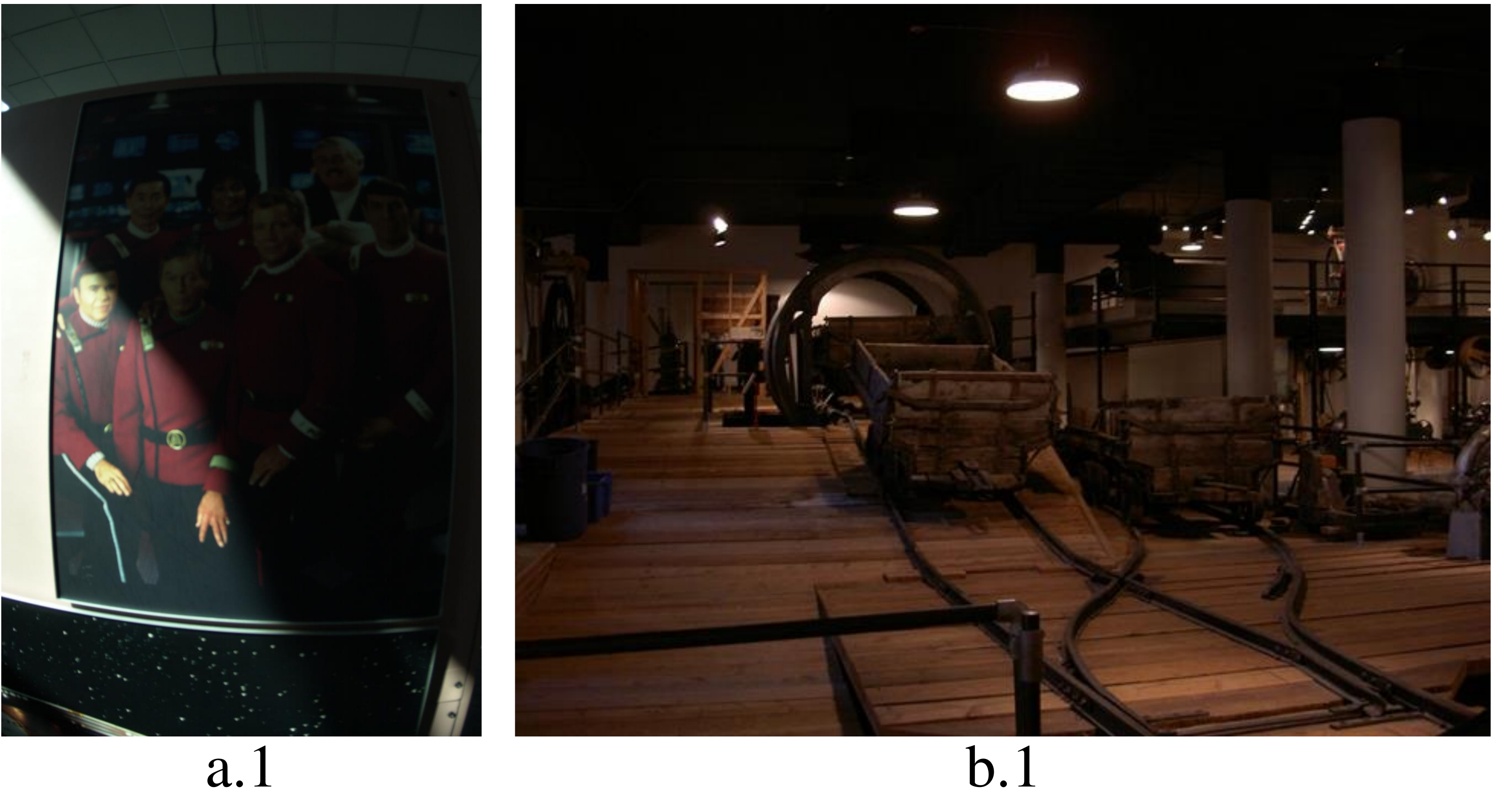}
	\caption{The different visual perceptions with various low-light images.}
\end{figure}
\begin{figure*}[t]
	\centering
	\includegraphics[width=1\linewidth]{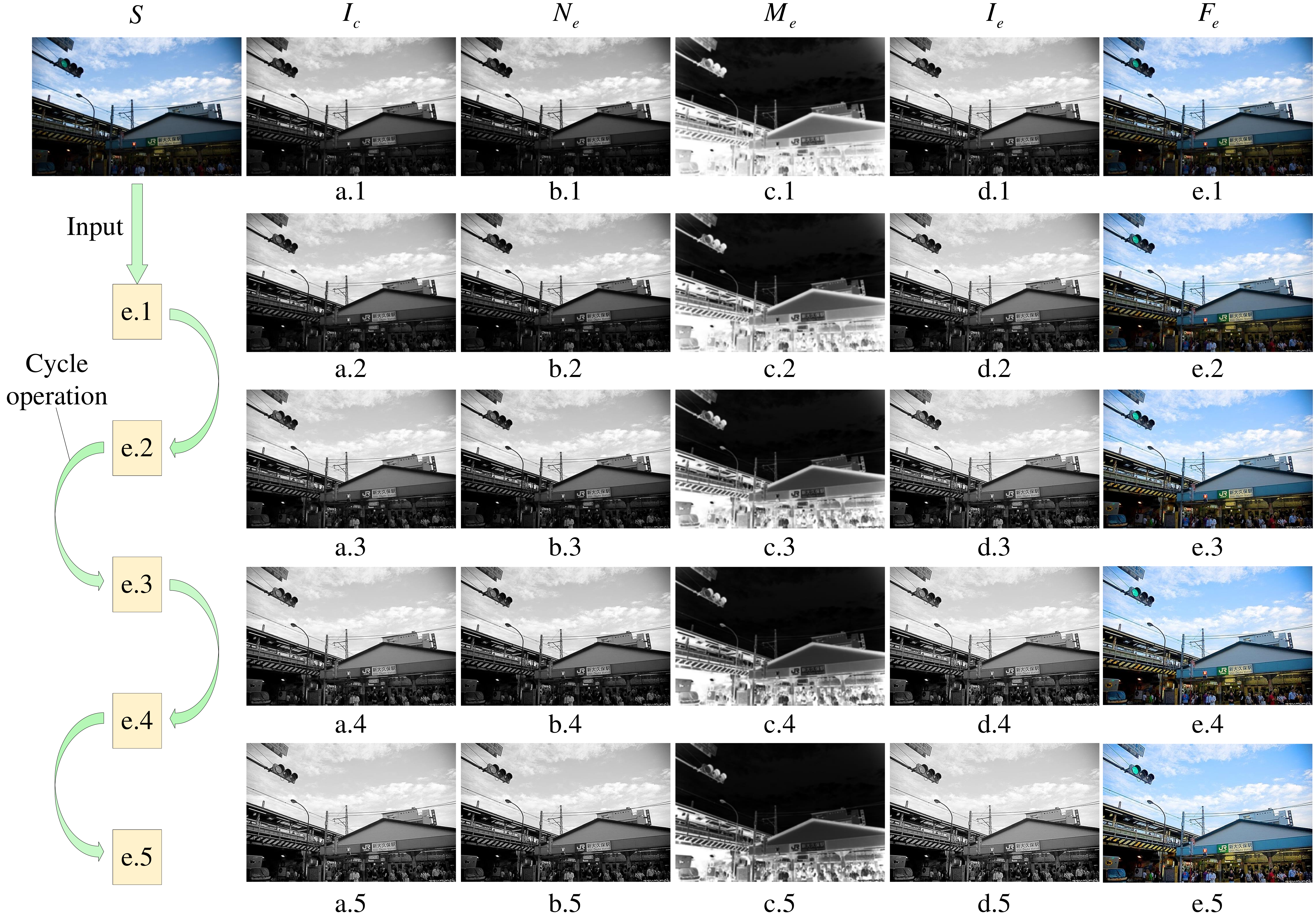}
	\caption{Illustration of the cycle operation for reconstructing brightness channel.}
\end{figure*}

\begin{figure}[t]
	\vspace{0cm}  
	\setlength{\abovecaptionskip}{0.2cm}   
	\setlength{\belowcaptionskip}{-0.5cm}   
	\centering
	\includegraphics[width=1\linewidth]{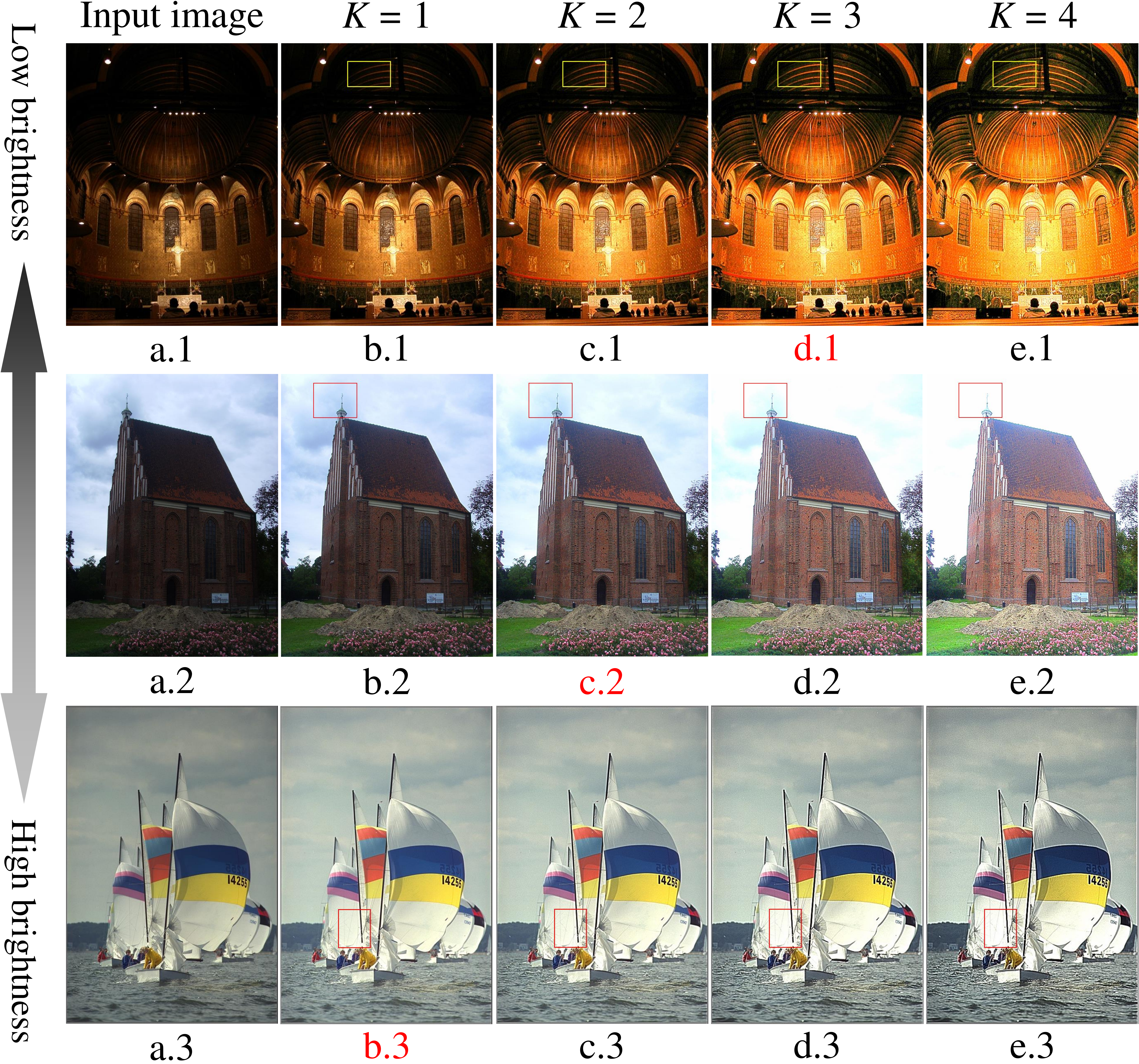}
	\caption{The enhanced results with the number of cycles $K$=1, 2, 3, 4. The \emph{optimal} results selected by the comparator are {\color{red}{d.1}}, {\color{red}{c.2}}, and {\color{red}{b.3}}.}
\end{figure}
\begin{figure*}[t]
	\setlength{\abovecaptionskip}{-0.1cm}
	\setlength{\belowcaptionskip}{-0.5cm}
	\centering
	\includegraphics[width=1\textwidth]{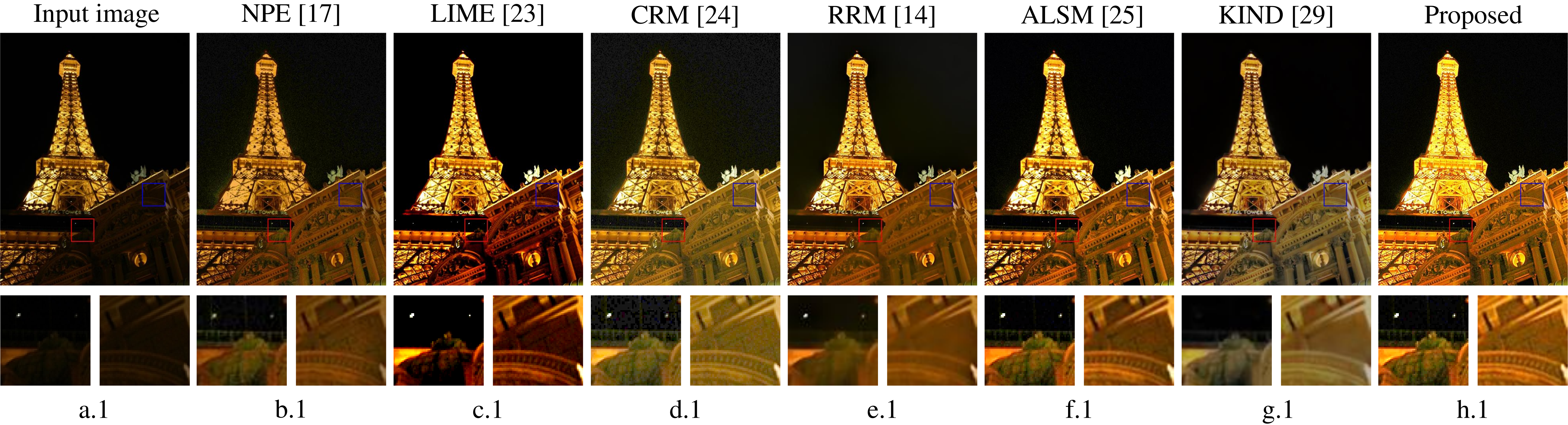}
	\caption{Adaptive ability of brightness enhancement comparison between the proposed method and the state-of-the-art methods. Test image is from DICM dataset \cite{ying2017new}.}
	\vspace{-0.4cm}
\end{figure*}
\begin{figure*}[t]
	\setlength{\abovecaptionskip}{-0.1cm}
	\setlength{\belowcaptionskip}{-0.5cm}
	\centering
	\includegraphics[width=1\textwidth]{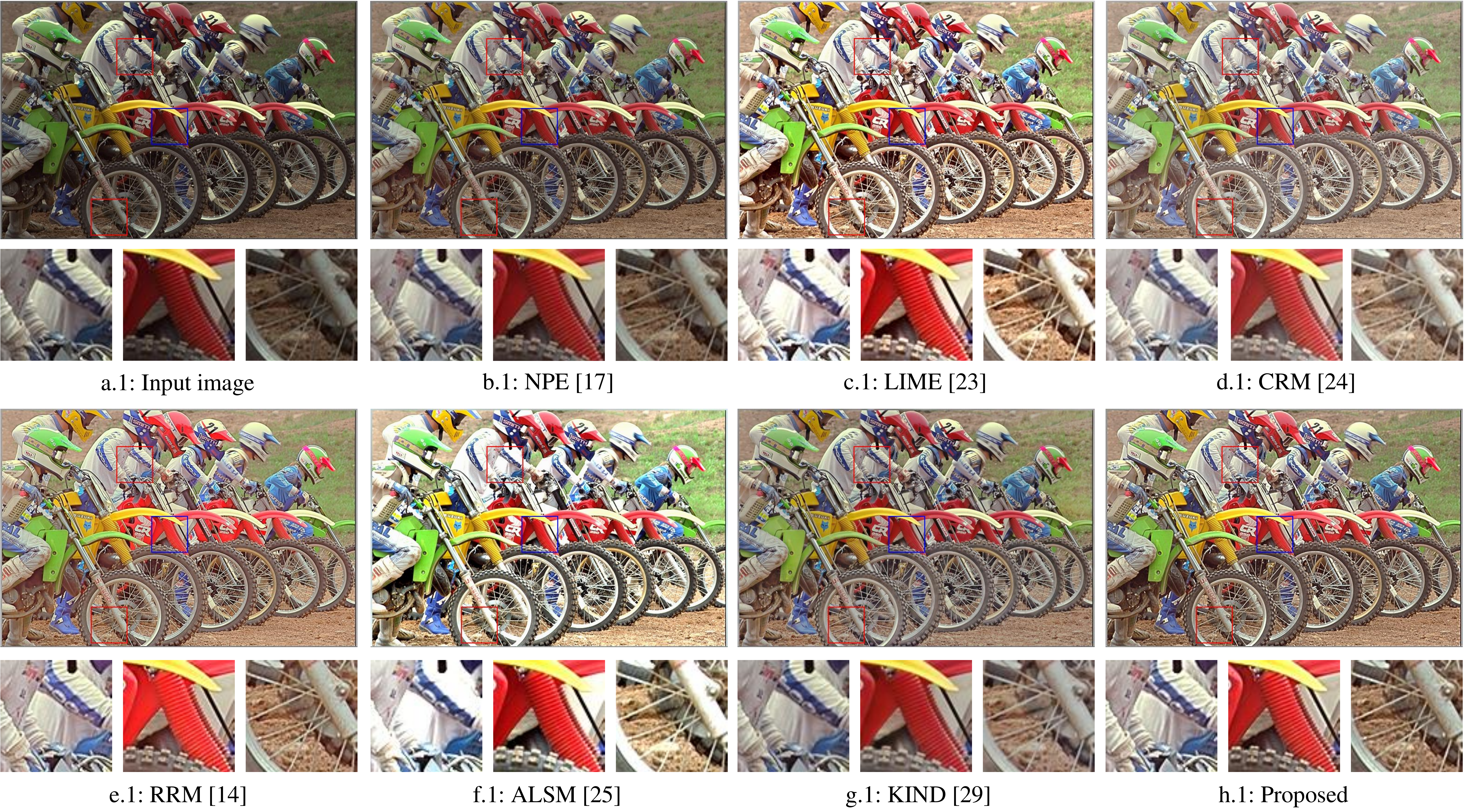}
	\caption{Adaptive ability of brightness enhancement comparison between the proposed method and the state-of-the-art methods. Test image is from Kodak dataset.}
	\vspace{-0.4cm}
\end{figure*}
\begin{figure*}[t]
	\vspace{-0.8cm}
	\setlength{\abovecaptionskip}{0.2cm}   
	\setlength{\belowcaptionskip}{-0.3cm}   
	\centering
	\includegraphics[width=1\textwidth]{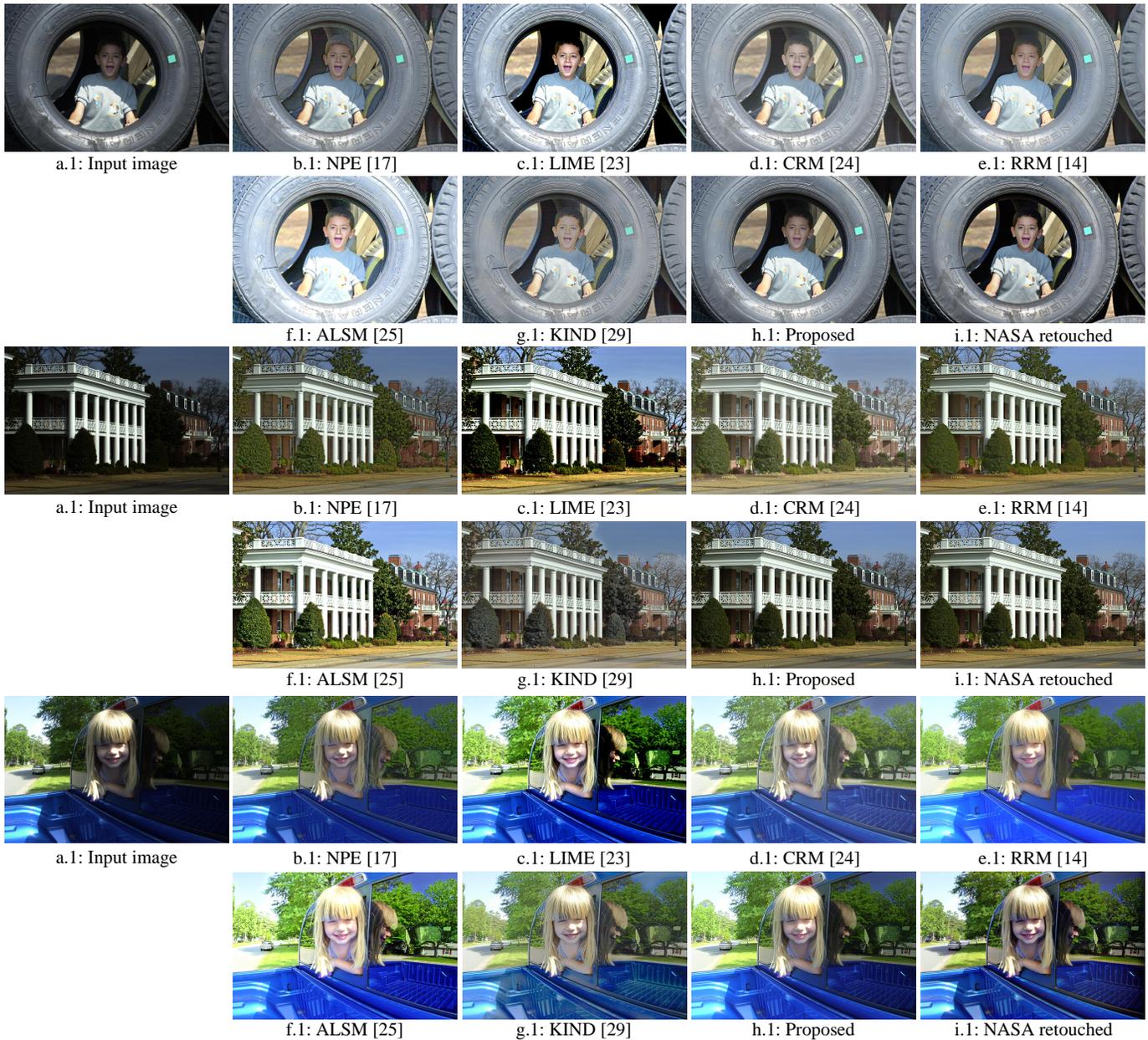}
	\caption{Visual comparison between the proposed method and the state-of-the-art methods. Test images are from Retinex dataset.}
\end{figure*}
\begin{figure*}[p]
	\vspace{0cm}
	\setlength{\abovecaptionskip}{0.2cm}   
	\setlength{\belowcaptionskip}{-0.4cm}   
	\centering
	\includegraphics[width=1\textwidth]{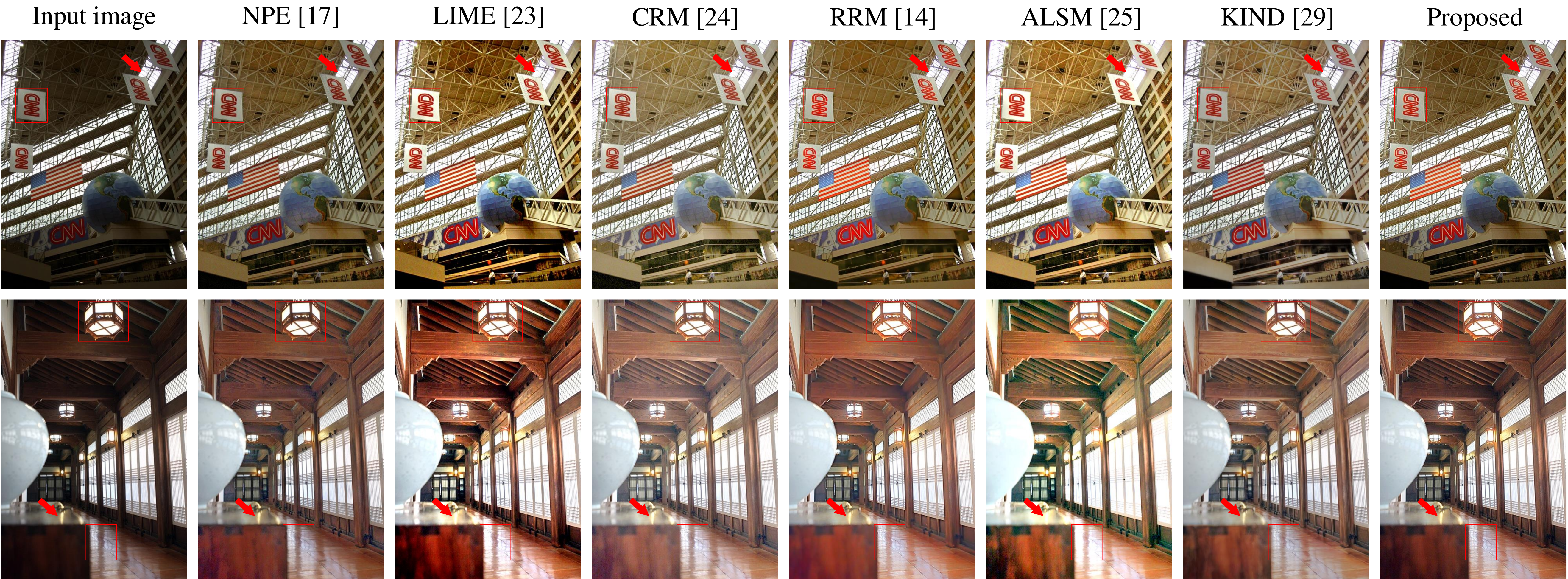}
	\caption{Visual comparison between the proposed method and the state-of-the art methods on DICM dataset.}
\end{figure*}
\begin{figure*}[p]
	\vspace{0cm}
	\setlength{\abovecaptionskip}{0.2cm}   
	\setlength{\belowcaptionskip}{-0.4cm}   
	\centering
	\includegraphics[width=1\textwidth]{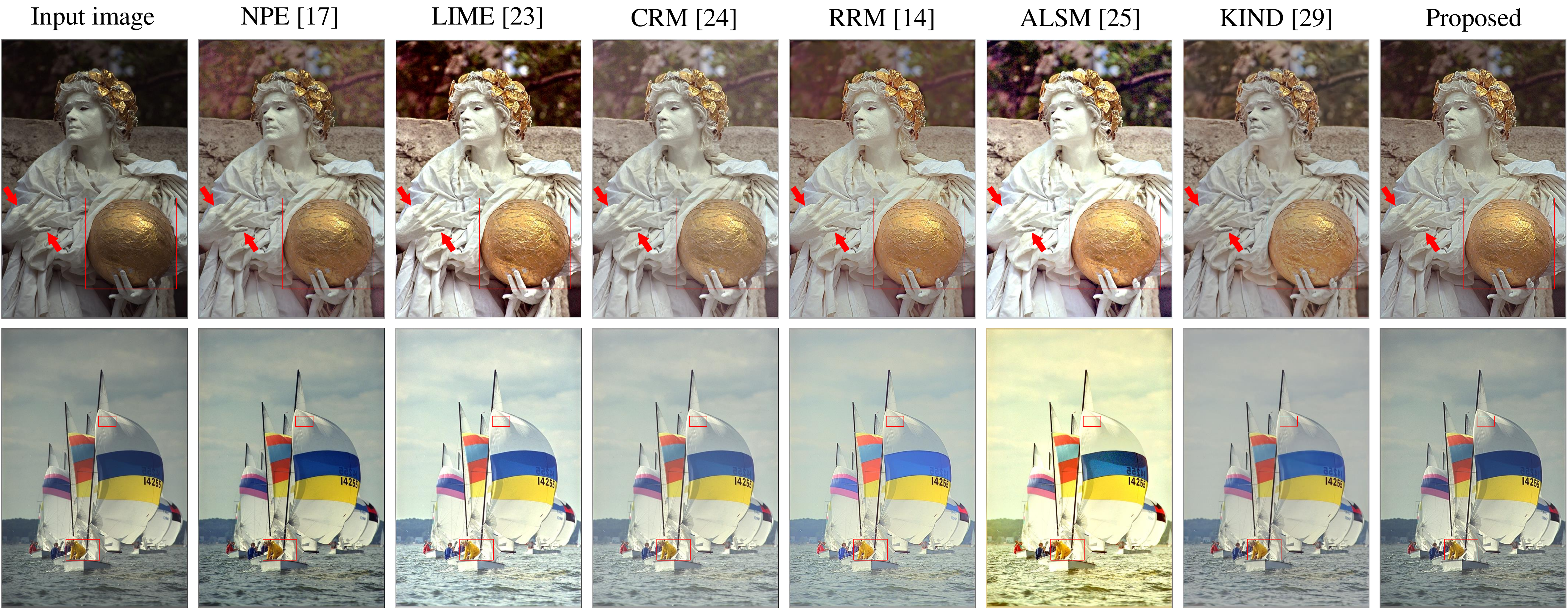}
	\caption{Visual comparison between the proposed method and the state-of-the art methods on Kodak dataset.}
\end{figure*}
\begin{figure*}[p]
	\vspace{0cm}
	\setlength{\abovecaptionskip}{0.2cm}   
	\setlength{\belowcaptionskip}{-0.4cm}   
	\centering
	\includegraphics[width=1\textwidth]{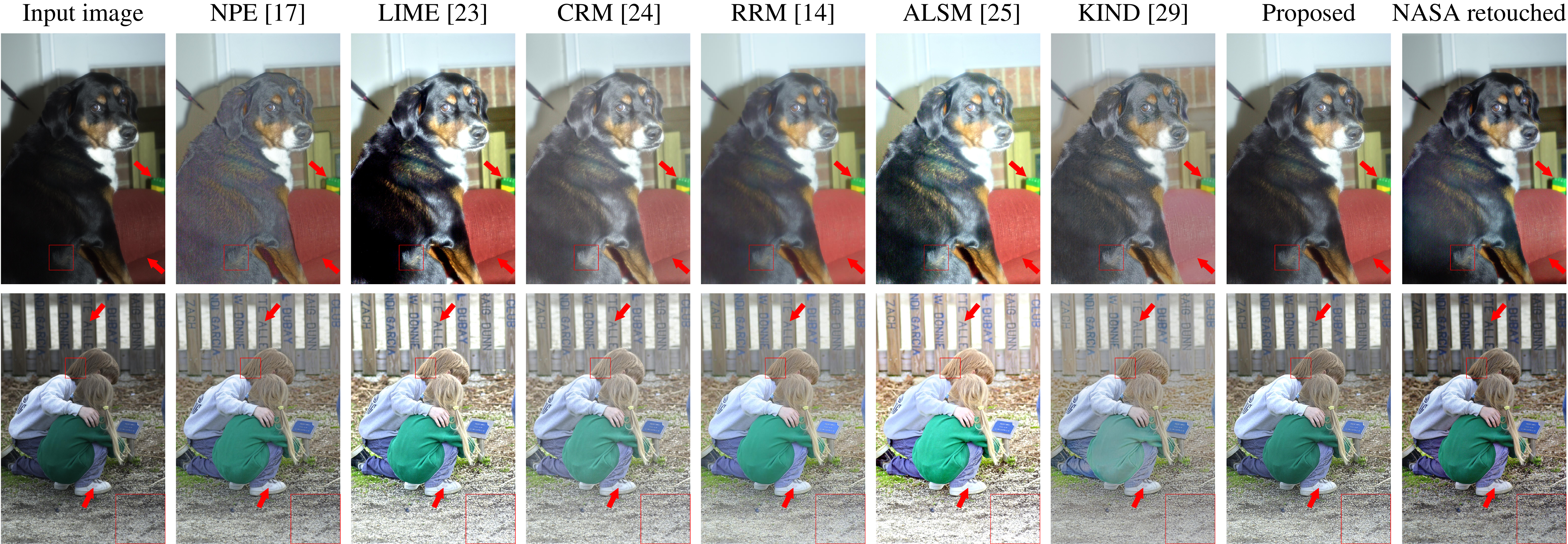}
	\caption{Visual comparison between the proposed method and the state-of-the art methods on Retinex dataset.}
\end{figure*}
\begin{figure*}[t]
	\centering
	\includegraphics[width=1\textwidth]{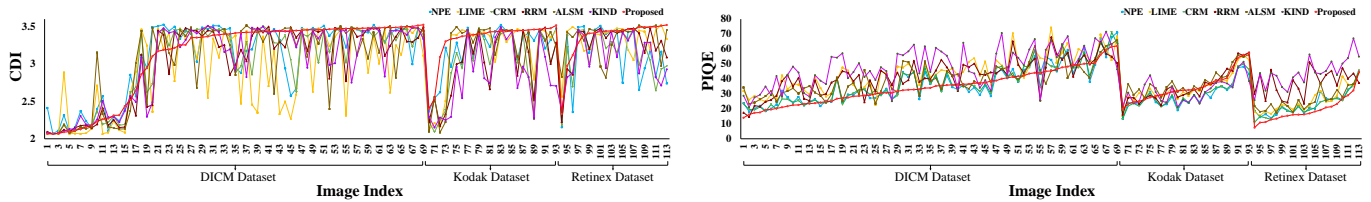}
	\caption{Quantitative result of each test example on DICM, Kodak, and Retinex dataset in terms of two no-reference evaluation metrics CDI and PIQE.}
\end{figure*}
\begin{figure*}[t]
	\centering
	\includegraphics[width=1\textwidth]{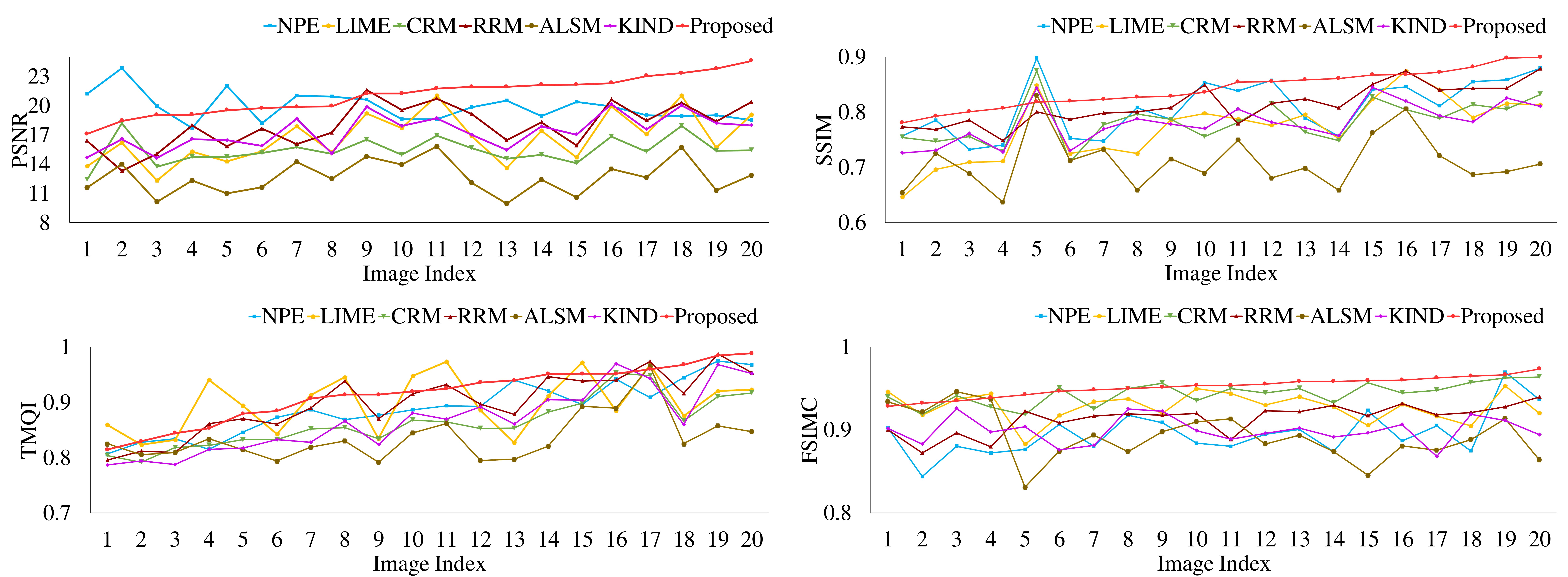}
	\caption{Quantitative result of each test example on Retinex dataset in terms of four full-reference evaluation metrics PSNR, SSIM, TMQI, and FSIMC.}
\end{figure*}

\subsection{VP Information Modeling}
The VP model aims to quantify human visual perception of low-light images with different feature information. Let's start with an observation of the characteristics of image information in Fig.4-a.1 and Fig.4-b.1. The lower left areas of Fig.4-a.1 is brighter than the upper right, and most of the image areas have low brightness in Fig.4-b.1. This indicates that vision is sensitive not only to the light intensity, but also to the distribution ratio of bright/dark areas. Based on this obervation, we propose VP to model interesting information. The mathematical quantification process is as follows:
\begin{eqnarray}
P_\alpha &=& \frac{\bm{\phi}_\alpha}{X\cdot{Y}-\nabla{v}} \, , 
\end{eqnarray}
\begin{eqnarray}
Q_\alpha &=& \frac{\bm{\varphi}_\alpha}{X\cdot{Y}-\nabla{v}} \, ,
\end{eqnarray}

\begin{eqnarray}
\phi_\alpha = \begin{cases} 
\sum\limits_{i=1}^{i=X}{\sum\limits_{j=1}^{j=Y} {H_b(i,j)}} \, ,  & \mbox{if }\alpha = 1 \, , \vspace{2.5ex}\\ 
\sum\limits_{i=1}^{i=X}{\sum\limits_{j=1}^{j=Y} {H_d(i,j)}} \, ,  & \mbox{if }\alpha = 2 \, , 
\end{cases} 
\end{eqnarray}
\begin{eqnarray}
\varphi_\alpha = \begin{cases} 
\left \{ m_1 = \begin{cases}
1, \:  \mbox{if }H_b > 0\\
0, \:  \mbox{if }H_b \leq 0\\
\end{cases} \! \left | \: \sum\limits_{i=1}^{i=X}{\sum\limits_{j=1}^{j=Y} {m_1(i,j)}} \right . \right \} \, , \\ \hspace*{4.78cm}
\enspace \mbox{if }\alpha = 1 \, , \vspace{3ex}\\ 
 \left \{ m_2 = \begin{cases}
1, \:  \mbox{if }H_d > 0\\
0, \:  \mbox{if }H_d \leq 0\\
\end{cases} \! \left | \: \sum\limits_{i=1}^{i=X}{\sum\limits_{j=1}^{j=Y} {m_2(i,j)}} \right . \right \} \, ,\\
\hspace*{4.78cm}
\enspace \mbox{if }\alpha = 2 \, , \\ 
\end{cases} 
\end{eqnarray}
where $P_\alpha$ represents bright or dark intensity. $Q_\alpha$ represents the area ratio of bright or dark areas. $\alpha$ denotes the decomposition scale of VP. $\alpha=[1,2]$ are the bright and dark areas respectively. $\phi_\alpha$ represents the sum of pixel values from bright or dark areas. $\varphi_\alpha$ represents the total number of pixels from bright or dark areas. $X$ and $Y$ is the image size. $\nabla{v}$ is the interference term, representing the total number of zero pixels in image.
\subsection{Adaptive Estimation}
\subsubsection{Visual Perception Quantification}
In Fig.3, the VP modeling visual information is prone to produce more eigenvalues, bringing high computation complexity. We therefore propose the $\bm\beta$ function to represent the degree of VPI. Since $P_2$ and $Q_2$ is the visual characteristics of dark areas, we define the quantization function of VPI as:
\begin{eqnarray}
\bm\beta &=& \sqrt{\exp{(\sqrt{\log(\frac{1}{P_2})}-\log(\frac{1}{Q_2}))}} \, .
\end{eqnarray}

In addition, since the input signals of equation (7) are all from the dark areas, $\bm\beta$ could not fully estimate the visual perception of bright areas. In order to remedy this deficiency, we design the $\bm\gamma$ function to assist $\bm\beta$ in estimating the reflectance. Therefore, $\bm\gamma$ has two roles: describe the visual perception information of the bright areas and assist $\bm\beta$ to estimate reflectance. $\bm\gamma$ is defined as:
\begin{eqnarray}
\bm\gamma = \begin{cases} 
P_1,  & \mbox{if }Q_1 > Q_2  \, , \vspace{1ex}\\
P_2, & \mbox{if }Q_1 \leq Q_2 \, .
\end{cases}
\end{eqnarray}

Equation (8) indicates that the $\bm\gamma$ obtains value by comparing the distribution ratio of bright/dark areas. When $Q_1 \leq Q_2$, the low-light image shows a darker state. For normalized images, $\bm\gamma$  with the smaller $P_2$ can obtain a larger enhancement intensity by exponentiation. Conversely, when $Q_1>Q_2$, a larger $\bm\gamma=P_1$ reduces the enhancement intensity of bright areas. Hence, this design could further improve the generalization of the proposed framework.

\subsubsection{Illumination Estimation}
Due to the obvious imbalance of visual photosensitivity, the enhancement weights with dark areas should be larger than that in bright areas. Therefore, we design two illumination regulators. The first one has the ability to quickly increase illumination in dark areas while the second one contributes to reduce over-saturation in the bright areas while enhancing illumination. The first regulator is described as:
\begin{eqnarray}
A &=& \lambda\big[\, \underbrace{(1+\log(1+\exp(\max(I_b)-I_b)))^4}_{\text{Enhancement Weights}} \, \cdot{I_b}\big] \, .
\end{eqnarray}

We take the output of the first regulator as the input signal of the second regulator to enhance the illumination in bright areas. This design could effectively avoid the illumination artifacts caused by separate operation. So the second regulator is described as:
\begin{eqnarray}
M_e &=& 2- {\left\lVert - \log(10-9 \frac{\max(A)-A}{\max(A)}) \right\lVert \,}^{\bm\beta^2} + \nabla{\theta}  \, , 
\end{eqnarray}
\begin{eqnarray}
\nabla{\theta} &=& \left\lVert \max{(A)} - \max{(A^{\,\bm\beta})} \right\lVert \, ,
\end{eqnarray}
where $\nabla{\theta}$ represents the correction factor, which adjusts the enhancement weights of bright areas. $\lVert\bullet\lVert$ is the absolute value sign.

\subsubsection{Reflectance Estimation}
In Section III-B, we indicate that the preprocessing process provides a high enhancement weights for the detail area. In this section, considering that the brightness channel usually contains rich image details, we utilize $I_c$ as the input image of reflectance estimation to obtain clear enhancement results. Reflectance estimation equation is as follows:
\begin{eqnarray}
N_e &=& I_c^{ \, U} \, ,
\end{eqnarray}
\begin{eqnarray}
U &=& \bigg[{\frac{1}{X\cdot Y}} \cdot\sum\limits_{i=1}^{i=X}\sum\limits_{j=1}^{j=Y}\big[I_b^{\bm\beta}(i,j)+1\big]^{\bm\gamma}\bigg]^{\frac{1}{\bm\beta}} \, ,
\end{eqnarray}
where $U \ge 1$, $I_c\in[0,1]$. $U$ as the exponentiation of $I_c$ could effectively prevent over-saturation of bright areas. Meanwhile, for different input images, the updated $\bm\gamma$ and $\bm\beta$ are applied to equation (13) to effectively increase the robustness of the framework.
\begin{table}[tp]  
	\centering  
	\caption{Comparison Of The Number Of Cycle Operation $K$ And Adaptive Threshold $T$ To Determine The Optimal Result. The Output Condition Satisfies $K = T$.}
	\label{tab:performance_comparison}  
	\begin{tabular}{|c|c|ccccc|}  
		\hline  
		\multicolumn{1}{|c|}{\multirow{6}{*}{Input image}}&
		\multicolumn{1}{c|}{\multirow{2}{*}{a.1}}& \multirow{2}{*}{\makecell[c]{$K=$ \\ $T=$}} & \multirow{2}{*}{\makecell[c]{1\\4}} & \multirow{2}{*}{\makecell[c]{2\\3}} & \multirow{2}{*}{\makecell[c]{\color{red}3\\\color{red}3}} & \multirow{2}{*}{\makecell[c]{4\\2}} \\
		&\multicolumn{1}{c|}{\multirow{2}{*}{}}  & \multirow{2}{*}{\makecell[c]{ \\ }} & \multirow{2}{*}{\makecell[c]{\\}}& \multirow{2}{*}{\makecell[c]{\\}}& \multirow{2}{*}{\makecell[c]{\\}}& \multirow{2}{*}{\makecell[c]{\\}} \\
		\cline{2-7}
		&\multicolumn{1}{c|}{\multirow{2}{*}{a.2}}  & \multirow{2}{*}{\makecell[c]{$K=$ \\ $T=$}}  & \multirow{2}{*}{\makecell[c]{1\\3}} & \multirow{2}{*}{\makecell[c]{\color{red}2\\\color{red}2}} & \multirow{2}{*}{\makecell[c]{3\\2}} & \multirow{2}{*}{\makecell[c]{4\\2}} \\
		&\multicolumn{1}{c|}{\multirow{2}{*}{}}& \multirow{2}{*}{\makecell[c]{ \\ }} & \multirow{2}{*}{\makecell[c]{\\}}& \multirow{2}{*}{\makecell[c]{\\}}& \multirow{2}{*}{\makecell[c]{\\}}& \multirow{2}{*}{\makecell[c]{\\}} \\
		\cline{2-7} 
		&\multicolumn{1}{c|}{\multirow{2}{*}{a.2}}& \multirow{2}{*}{\makecell[c]{$K=$ \\ $T=$}}  & \multirow{2}{*}{\makecell[c]{\color{red}1\\\color{red}1}} & \multirow{2}{*}{\makecell[c]{2\\1}} & \multirow{2}{*}{\makecell[c]{3\\1}} & \multirow{2}{*}{\makecell[c]{4\\1}} \\
		&\multicolumn{1}{c|}{\multirow{2}{*}{}}& \multirow{2}{*}{\makecell[c]{ \\ }} & \multirow{2}{*}{\makecell[c]{\\}}& \multirow{2}{*}{\makecell[c]{\\}} & \multirow{2}{*}{\makecell[c]{\\}}& \multirow{2}{*}{\makecell[c]{\\}} \\
		\hline  
	\end{tabular}  
\end{table}
\subsection{Image Reconstruction}
We reconstruct the enhanced result in three main steps. First, as shown in Fig.3, the new brightness channel is constructed by using reflectance and illumination. Then, we reconstruct the intermediate image $F_e$ by converting of HIS to RGB. Finally, since the cycle operation can produce multiple intermediate images, we set a comparator to select the optimal output. This process is given as follows:

Step 1: brightness channel reconstruction by Redinex model.
\begin{eqnarray}
I_e &=& {N_e}\otimes{M_e} \, .
\end{eqnarray}

Step 2: space conversion by HIS to RGB.
\begin{eqnarray}
F_e &=& {{\rm HSI}(I_e)} \to {\rm RGB} \, ,
\end{eqnarray}
where $HSI(I_e)$: HSI color space with a enhanced brightness channel. $\to$: transform the relationship. RGB: RGB color space.

Step 3: output using the comparator.
\begin{eqnarray}
F_o &=& F_e\, , \,\,\,\,\, \mbox{if }K = T = \lfloor \, (\bm\beta^2)^{\sqrt{\bm\beta}} \, \rfloor \, ,
\end{eqnarray}
where $K$ is the number of cycle operation. $T$ represents adaptive threshold. $\lfloor\bullet\rfloor$ is rounding operation. We will analyze the comparator detailedly in Section IV-B.

In generally, we summarize the establishment of the proposed framework as follows. First, we implement the preprocessing to improve the accuracy of VP model as well as the image details. Second, we propose the VP model to precisely simulate the process of HVS perceiving image information. Third, according to the quantified $\bm\beta$ function, we design an adaptive scheme to estimate illumination and reflectance. Finally, we design a optimal determination strategy, consisting of a cycle operation and a comparator to give the optimal reconstructed enhanced image.

\section{Experiment}
\subsection{Experiment Setting}
\textbf{Baseline}. In this section, we choose six state-of-the-art methods, including KIND (“kindling the darkness” on deep learning) \cite{zhang2019kindling}, ALSM (using absorption light scattering model) \cite{wang2019low}, RRM (using robust Retinex model) \cite{li2018structure}, CRM (using camera response model) \cite{ying2017new}, LIME (using a method for low-light image enhancement) \cite{guo2016lime2}, and NPE (using naturalness preserved enhancement algorithm) \cite{wang2013naturalness} as the baseline to verify the effectiveness of the proposed framework. 

\textbf{Benchmark and implementation platform}. We evaluate our framework on three benchmark datasets, including Retinex\footnote{https://dragon.larc.nasa.gov/retinex/pao/news/}, Kodak\footnote{http://r0k.us/graphics/kodak/}, and DICM \cite{ying2017new}. Retinex dataset provides 20 ground truth enhanced images retouched by National Aeronautics and Space Administration (NASA). Kodak dataset contains 24 lossless true color images. DICM dataset contains 69 low-light images from commercial digital cameras. In addition, we download highly exposed images from Internet to further test the adaptability of proposed framework. All experiments are implementation on MATLAB. 

\textbf{Evaluation metrics}. Since Kodak and DICM dataset do not provide ground truth enhanced images, no-reference evaluation metrics are necessary. For the Retinex dataset with ground truth enhanced images, we use both the no-reference and full-reference evaluation metrics to measure the performance of various methods. Also, considering a single evaluation metric may not reflect the enhancement performance, we harness multiple evaluation metrics together, including: CDI (contrast distorted images assessment) \cite{fang2014no}, PIQE (perception based image quality evaluator) \cite{venkatanath2015blind}, PSNR (peak signal to noise ratio, provided by MATLAB) and SSIM (structural similarity index) \cite{wang2004image}, TMQI (tone-mapped image quality index) \cite{yeganeh2012objective}, and FSIMC (feature similarity index for color image) \cite{zhang2011fsim}. Specifically, we harness two no-reference evaluation metrics, CDI and PIQE to conduct evaluation on three datasets, respectively; then we use four full-reference evaluation metrics, PSNR, SSIM, TMQI, and FSIMC to employ evaluation on Retinex dataset. Note that a larger value of each evaluation metric indicates a better enahcenment result except for the PIQE, where a lower value gives a better performance. 
\subsection{Adaptability Analysis}
We analyze the proposed framework from two aspects, i.e., \emph{cycle operation} and \emph{optimal result selection} to verify its adaptive ability. In the cycle operation experiment, we discuss the adjustment process of image brightness information with the increase of the number of cycles. We then empirically demonstrate the effectiveness of the designed comparator.

\textbf{Cycle Operation}. In order to intuitively analyze the process of adaptive enhancement, we give an example in Fig.5 to exhibit the key results with five cycle operations, including: $I_c$ (brightness channel image), $N_e$ (estimated reflectance), $M_e$ (estimated illumination), $I_e$ (reconstructed brightness channel image), and $F_e$ (intermediate image). 

The input image is a challenging case with conspicuous contrast between bright and dark areas. In the process of image enhancement, the dark areas usually needs more enhancement intensity than the bright areas, otherwise it will cause local over-saturation. Hence, an adaptive enhancement framework should be able to automatically adjust the enhancement intensity by the HVS's perception in bright and dark areas. In Fig.5, we observe that the reconstructed $I_e$ improves the visual effect of the dark areas while effectively avoiding the over-saturation of the bright areas. This mainly benefits from the following two aspects: first, since the $N_e$ contains details from $I_c$,  the brightness channel $I_e$ reconstructed by $N_e$ and $M_e$ could effectively enhance the image details; second, the $M_e$ provides the another contribution from the two illumination regulators (see equation (9) and (10) ) that assign high and low enhancement weights for dark and light areas respectively. Furthermore, with the cycle operation, the reconstructed $I_e$ is gradually stabilized in Fig.5, which is because we embed $\beta$ into the illumination and reflectance estimation to increase the adaptability. Through the analyses above, we can see that the proposed method achieves the expected adaptive ability. However, more cyclc operations bring longer running time. To balance the enhancement result and computational efficiency, we utilize equation (16) as the optimal output condition to decide the number of cycle operation.

\begin{table}[t]
	\setlength{\abovecaptionskip}{0.1cm}   
	\centering
	\setlength{\tabcolsep}{0.9mm}{
		\caption{Average Quantitative Results of Various Methods on DICM Dataset (Including 69 Images) In Terms Of Two No-reference Evaluation Metrics.}	
		\begin{tabular}{cccccccc}
			\midrule
			Metric & NPE & LIME & CRM & RRM & ALSM & KIND & Proposed\\
			\midrule
			CDI  & \color{blue}3.0873 & 2.9021 & 3.0442 & 3.0171 & 3.0447 & 3.0378 & \color{red} 3.1068\\
			PIQE & \color{blue}37.1805 & 44.5177 & 37.3928 & 42.0823 & 39.9361 & 48.0682 & \color{red} 34.9037\\
			\midrule
	\end{tabular}}	
\end{table}
\begin{table}[t]
	\setlength{\abovecaptionskip}{0.1cm}   
	\centering
	\setlength{\tabcolsep}{0.9mm}{
		\caption{Average Quantitative Results of Various Methods on Kodak Dataset (Including 24 Images) In Terms Of Two No-reference Evaluation Metrics.}	
		\begin{tabular}{cccccccc}
			\midrule
			Metric & NPE & LIME & CRM & RRM & ALSM & KIND & Proposed\\
			\midrule
			CDI  & \color{blue}3.2113 & 3.1732 & 3.0269 & 3.0191 & 3.1241 & 2.8483 & \color{red} 3.3287\\
			PIQE & \color{red}29.0549 & 34.4885 & \color{blue}29.5458 & 34.7636 & 37.5778 & 32.0872 & 34.8659\\
			\midrule
	\end{tabular}}	
\end{table}
 \begin{table}[t]
 	\setlength{\abovecaptionskip}{0.1cm}   
 	\centering
 	\setlength{\tabcolsep}{0.9mm}{
 		\caption{Average Quantitative Results of Various Methods on Retinex Dataset (Including 20 Images) In Terms Of Two No-reference Evaluation Metrics And Four Full-reference Evaluation Metrics.}	
 		\begin{tabular}{cccccccc}
 			\midrule
 			Metric & NPE & LIME & CRM & RRM & ALSM & KIND & Proposed\\
 			\midrule
 			CDI   & 3.1213 & 3.3273 & 3.1295 & 3.2604 & \color{blue}3.3250 & 3.1159 & \color{red} 3.3535\\
 			PIQE  & 23.4588 & 25.8017 & \color{blue}22.6168 & 37.4695 & 26.1850 & 44.3455 & \color{red} 19.3787\\
 			PSNR  & \color{blue}19.8969 & 16.6846 & 15.4217 & 17.9710 & 12.6620 & 17.3198 & \color{red} 21.1191\\
 			SSIM  & 0.8097 & 0.7724 & 0.7834 & \color{blue}0.8139 & 0.7103 & 0.7808 & \color{red}0.8427\\
 			TMQI  & 0.8904 & 0.8990 & 0.8631 & \color{blue}0.8996 & 0.8363 & 0.8681 & \color{red}0.9161\\
 			FSIMC  & 0.8958 & 0.9279 & \color{blue}0.9437 & 0.9135 & 0.8925 & 0.8994 & \color{red} 0.9520\\
 			\midrule
 	\end{tabular}}	
 \end{table}
 \begin{table}[t]
 	\setlength{\abovecaptionskip}{0.1cm}   
 	\centering
 	\setlength{\tabcolsep}{0.9mm}{
 		\caption{Average Quantitative Results of Various Methods on DICM, Kodak, Retinex Datasets (113 Images in Total) In Terms Of Two No-reference Evaluation Metrics.}	
 		\begin{tabular}{cccccccc}
 			\midrule
 			Metric & NPE & LIME & CRM & RRM & ALSM & KIND & Proposed\\
 			\midrule
 			CDI  & \color{blue}3.1197 & 3.0349 & 3.0556 & 3.0606 & 3.1112 & 3.0113 & \color{red}3.1976\\
 			PIQE & \color{blue}33.0261 & 39.0751 & 33.1110 & 39.7115 & 37.0014 & 44.0151 & \color{red}32.1479\\
 			\midrule
 	\end{tabular}}
 \end{table}
 \begin{figure*}[t]
 	\vspace{-0.8cm}
 	\setlength{\abovecaptionskip}{0.2cm}   
 	\setlength{\belowcaptionskip}{-0.3cm}   
 	\centering
 	\includegraphics[width=1\textwidth]{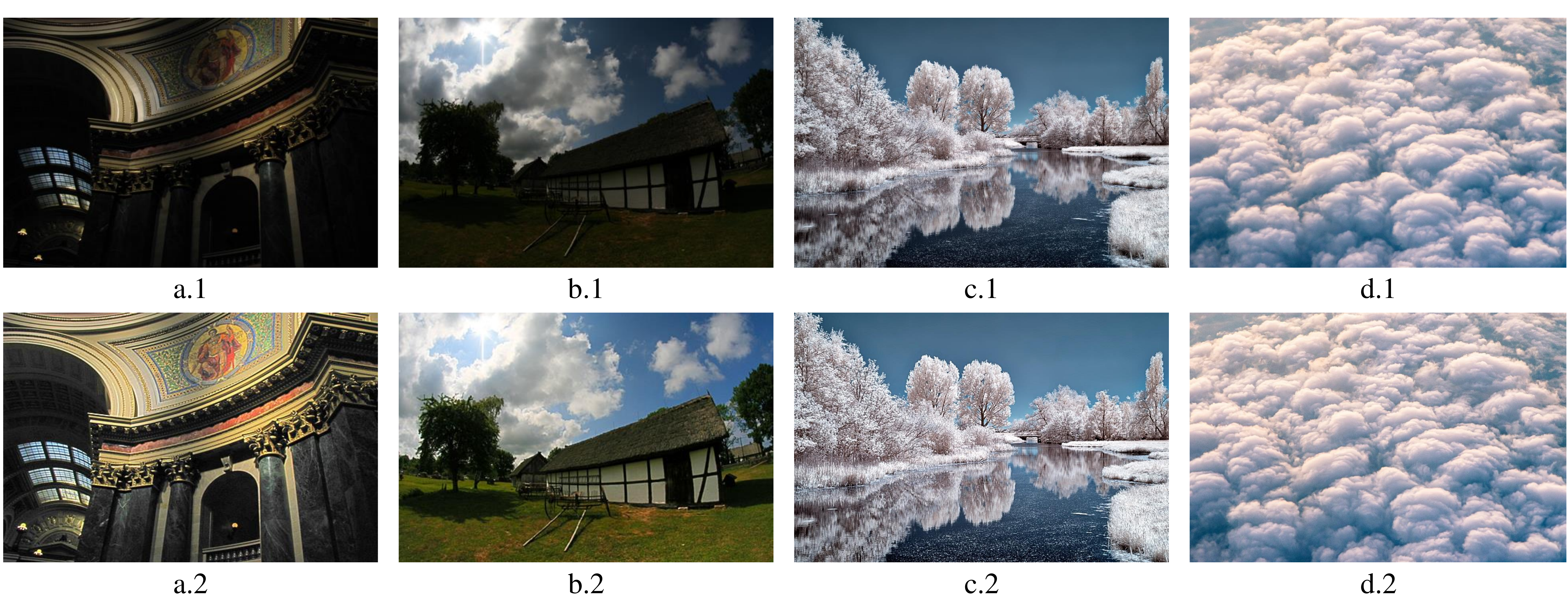}
 	\caption{The enhancement results of various challenging cases using our method.}
 \end{figure*}
 \begin{table*}[h]
 	\setlength{\abovecaptionskip}{0.1cm}   
 	\centering
 	\caption{Computational Efficiency Comparison Between The Proposed Method and The State-of-The Art Methods. (Unit: second).}
 	\begin{tabular}{ccccccccc}
 		\midrule
 		Size (width$\times$height) & 100$\times$200 & 200$\times$250 & 250$\times$300 & 300$\times$300 & 350$\times$450 & 400$\times$500 & 500$\times$800 & 500$\times$1000\\
 		\midrule
 		NPE  & 0.5239 & 1.3296 & 1.8984 & 2.3389 & 4.0101 & 5.1531 & 10.2472 & 12.7174\\
 		LIME  & 0.1966 & 0.2736 & 0.3356 & 0.4343 & 0.8268 & 1.0613 & 2.0725 & 2.6115\\
 		CRM  &  \color{blue}0.0223 & \color{blue} 0.0639 & \color{blue} 0.0739 & \color{blue} 0.0761 & \color{blue} 0.1284 & \color{blue} 0.1704 & \color{blue} 0.2803 & \color{blue} 0.3383\\
 		RRM  & 0.5535 & 1.6979 & 2.5798 & 3.2416 & 4.8236 & 6.6187 & 16.4253 & 19.9932\\
 		ALSM  & 1.0169 & 2.2139 & 2.8572 & 3.7347 & 6.9059 & 8.6503 & 17.1045 & 21.9654\\
 		KIND  &1.6905 & 1.8361 & 1.8914 & 1.9142 & 2.1349 & 2.2331 & 2.7789 & 3.0585 \\
 		Proposed ($K$=1) & \color{red} 0.0179 & \color{red} 0.0276 & \color{red} 0.0393 & \color{red} 0.0425 & \color{red} 0.0885 & \color{red} 0.1022 & \color{red} 0.2135 & \color{red} 0.3176\\
 		Proposed ($K$=2) & 0.0244 & 0.0642 & 0.0851 & 0.1031 & 0.1749 & 0.2689 & 0.4617 & 0.5682\\
 		Proposed ($K$=3) & 0.0414 & 0.0801 & 0.1048 & 0.1242 & 0.2311 & 0.2801 & 0.5872 & 0.7653\\
 		\midrule
 	\end{tabular}
 \end{table*}

\textbf{Optimal result selection}. In this section, we discuss the comparator's (on equation (16)) ability to adaptively select optimal results from multiple cycle operations. We choose three input images with different brightness in Fig.6-a.1-a.2-a.3. Fig.6 shows various enhancement results with cycle number $K=1,2,3,4$, respectively.

We can observe that the smaller $K$ corresponds to the low brightness result, whereas the high brightness result means the larger $K$. In detail, the Fig.6-b.1-c.1-b.2 show the enhancement result of underexposure. Conversely, the Fig.6-e.1-d.2-e.2-c.3-d.3-e.3 expose the results of local overexposure. Therefore, according to visual analysis, the optimal results are Fig.6-d.1, Fig.6-c.2, and Fig.6-b.3. In order to accomplish this task adaptively, we set a threshold $T$ to select the optimal result in TABLE II. It is worth noting that the results of TABLE II and Fig.6 are based on the same input image. The red marks with $K =T$, in TABLE II and Fig.6, have the same number of cycles $K$, which indicates that the comparator could adaptively select the optimal output and improve the robustness of the proposed framework.

\subsection{Adaptability Comparison}
In Section IV-B, we demonstrate the calculation process of the proposed framework and emphasize the theoretical feasibility of adaptive image enhancement. In this section, we further compare the adaptive ability of the proposed method with ohter state-of-the-art methods. We evaluate the proposed method from two aspects, namely image brightness and image detail. 

\textbf{Image brightness comparison}. In Fig.7, we give an example of severe brightness distortion to compare the adaptability of image brightness enhancement. The test image is a challenging case, due to uneven light and imperceptible details. We can see that NPE and LIME exhibit insufficient brightness enhancement. Also, CRM, RRM, ALSM and KIND reveal varying degrees of detail distortion. In particular, RRM and KIND produce almost invisible details due to the weak brightness enhancement ability. By contrast, our method is more adaptive for enhancing images with severe brightness distortion.

\textbf{Image detail comparison}. In Fig.8 we show an example that contains rich image details to compare the adaptability of image detail enhancement. This example we test provides three local zoomed view areas with diverse image element, including: “clothing", “plastic", and “wheels". It can be observed from Fig.8 that NPE shows dim detail information with low brightness in Fig.8-b.1. In addition, CRM, RRM and KIND provide blurred image results due to low contrast (e.g. the “plastic" and “wheels" in Fig.8-d.1-e.1-g.1). LIME and ALSM lose image details due to the over-saturation problem (e.g. the “clothing" and “wheels" in Fig.8-c.1-f.1). By contrast, the proposed method exhibits more details in all the zoomed view areas, which further demonstrates that our method has stronger adaptability in enhancing different types of image details.

\subsection{Evaluation And Discussion}
\textbf{Visual comparison}. We first evaluate the proposed method from visual comparison. Fig.9 shows three representative visual comparison examples where we can observe that all the comparison methods exhibit the unnatural enhanced effect. For instance, NPE shows an low-brightness enhancement result; also, it produces obvious noises in dark areas (in Fig.9-b.1-b.2). LIME obtains high contrast images by sacrificing the local brightness (in Fig.9-c.3) and visually presenting an unreal oil painting effect (in Fig.9-c.1-c.2). CRM, RRM, and KIND produce atomization effect, resulting in unreal brightness enhancement (e.g. Fig.9-d.1-d.2-e.1-e.3-g.1-g.3). Further, the local color distortion leads to a poor visual perception (in Fig.9-d.3-g.1). ALSM loses details and textures (in Fig.9-f.1-f.2-f.3). Compare the enhancement results obtained by our method with that of others, we can observe that the proposed method achieves more natural enhancement results, balancing well various visual information, including: brightness, texture, and color; besides, our result is closest to the NASA rendered ground truth enhanced images. Further, we also exhibit more visual comparisons in Fig.10, Fig.11 and Fig.12. These examples above demenstrate that our method achieves the best visual perception. 

\textbf{Quantitative assessment}. We then verify our method through quantitative assessment. We first show the quantitative result of each test example on DICM, Kodak, and Retinex dataset in terms of two no-reference evaluation metrics CDI and PIQE in Fig.13. Then, we employ four full-reference evaluation metrics PSNR, SSIM, TMQI, and FSIMC to test each example in Retinex dataset, and the results are given in Fig.14. Intuitively, we can observe from Fig.13 and Fig.14 that our method reaches a leading score trend in almost each test example, not only on each no-reference evaluation metric, but also each full-reference evaluation metric. Quantitatively, we further calculate the average results of the DICM, Kodak, Retinex dataset in terms of two no-reference evaluation metrics in TABLE III, TABLE IV, and TABLE V, respectively. The best and second best results are highligted in red and blue respectively. We can see that the proposed method achieves the best scores in almost all cases. We also list the average results of four full-reference evaluation metrics on Retinex dataset and the average results of two no-reference evaluation metrics on three datasets (113 images in total) in Table V and TABLE VI, respectively. It can be seen that our method achieves the best scores in terms of all the evaluation metrics, which is consistent with the visual comparison performance. To sum up, the superior performance of visual comparison and quantitative assessment indicate that the proposed method acquires more natural images and can effectively balances the image brightness, contrast, texture structure, and color, bringing a better visual perception for HVS and higher quality enhanced results for facilitating other vision tasks. 

\subsection{Computational Efficiency}
In this section, we show the computational efficiency of various methods. Since the running time of the proposed method is affected by the number of cycles $K$, we show the running time when $K$ =1, 2 and 3 in TABLE VII. We can observe that our method achieves the higest computational efficiency when $K$ =1. Even when $K$ =3, our method also has faster running time than some of the state-of-the-arts, revealing its potential applications in real scenarioes. 

\subsection{Challenging Case Enhancement}
We also show the performance of our method in dealing with the challenging cases in Fig.15. Fig.15-a.1 is an underexposed case that has invisible details on “windows” and “marble” areas. In the Fig.15-b.1, we exhibit an uneven exposure image where the “clouds" areas are partially exposed while the “grass" areas are underexposed. Moreover, images with extreme local over-saturaion in bright foreground , such as “snow” areas and “cloud"  areas is given in Fig.15-c.1-d.1. We can observe that our method effectively overcomes the problem of over-saturation and thus acquires natural visual perception effects. In summary, our method performs well in enhancing various challenging cases and can obtain reasonable enhancement results.

\section{Conclusion}

In this paper, we proposed the VP model to simulate the relationship between light source and HVS, aiming at quantifying the visual perception information of images. We then presented rapid and adaptive $\bm\beta$ and $\bm\gamma$ functions as the illumination and reflectance estimation scheme to adjust the enhancement weights. Finally, considering the different low-light images usually have distinct visual perception, we introduced a optimal determination strategy, being composed of a \emph{cycle operation} and a \emph{comparator} to select the \emph{optimal} enhancement results. By leveraging the proposed VP model, illumination and reflectance estimation scheme, and optimal determination strategy, we established a rapid and adaptive framework for low-light image enhancement. We verified its adaptability from two aspects of image brightness comparison and image detail comparison. We then evaluated our method from extensive experiments. Results demenstrated that our method outperformed the state-of-the-arts qualitatively and quantitatively, and had a higher computational efficiency. Also, the proposed method performed well in enhancing various challenging cases, such as \emph{underexposed}, \emph{uneven exposure}, and \emph{extreme local over-saturaion} conditions. The MATLAB implementation of our method will be available at: {\color{blue}{https://github.com/MDLW/Low-Light-Image-Enhancement}}.

\ifCLASSOPTIONcaptionsoff
\newpage
\fi

\bibliographystyle{IEEEtran}           
\bibliography{VP_model}
\end{document}